\documentclass[aps,prl,twocolumn,amsmath,10pt,superscriptaddress,floatfix,nofootinbib]{revtex4-1}

\usepackage{amsmath}
\usepackage{epsfig}
\usepackage{color,soul}
\usepackage{multirow}

\newcommand{\state}[4]{{^{#1}\hspace{-0.6mm}{#2}_{#3}^{[#4]}}}

\newcommand\CScSa{\state{3}{S}{1}{1}}
\newcommand\COaSz{\state{1}{S}{0}{8}}
\newcommand\COcSa{\state{3}{S}{1}{8}}
\newcommand\COcPz{\state{3}{P}{0}{8}}
\newcommand\COcPj{\state{3}{P}{J}{8}}

\newcommand\mo{{\mathcal O}}

\newcommand{\LDME}[2]{\langle\mo^{#1}(#2)\rangle}

\newcommand{\mopa}{\LDME{\jsi}{\COaSz}}

\newcommand\mopc{\LDME{\jsi}{\COcPz}}
\newcommand\mopj{\LDME{\jsi}{\COcPj}}

\newcommand{\LDMEn}[1]{\langle\mo^{#1}_{[n]}\rangle}
\newcommand\mopn{\LDMEn{\jsi}}

\def\bea#1\eea{\begin{align}#1\end{align}}
\newcommand{\nnu}{\nonumber\\}
\newcommand{\bef}{\begin{figure}[h!tb]\centering}
\newcommand{\eef}{\end{figure}}
\newcommand{\jsi}{J/\psi}
\newcommand{\ccb}{c\bar{c}}

\begin{document}
\title{Exploring $\jsi$ production mechanism at the future Electron-Ion Collider}
                   
\author{Jian-Wei Qiu}
\email{jqiu@jlab.org}
\affiliation{Theory Center, Jefferson Lab, 12000 Jefferson Avenue, Newport News, VA 23606, USA}
                                
\author{Xiang-Peng Wang}
\email{xiangpeng.wang@anl.gov}
\affiliation{High Energy Physics Division, Argonne National Laboratory, Argonne, Illinois 60439, USA}

\author{Hongxi Xing}
\email{hxing@m.scnu.edu.cn}
\affiliation{Guangdong Provincial Key Laboratory of Nuclear Science, Institute of Quantum Matter,
 South China Normal University, Guangzhou 510006, China}

\date{\today}         

\begin{abstract}
We propose to use transverse momentum $p_T$ distribution of $\jsi$ production at the future Electron Ion Collider (EIC) to explore the production mechanism of heavy quarkonia in high energy collisions.  
We apply QCD and QED collinear factorization to the production of a $\ccb$ pair at high $p_T$, and non-relativistic QCD factorization to the hadronization of the pair to a $\jsi$.  We evaluate $\jsi$ $p_T$-distribution at both leading and next-to-leading order in strong coupling, and show that production rates for various color-spin channels of a $\ccb$ pair in electron-hadron collisions are very different from that in hadron-hadron collisions, which provides a strong discriminative power to determine various transition rates for the pair to become a $\jsi$.  We predict that the $\jsi$ produced in electron-hadron collisions is likely unpolarized, and the production is an ideal probe for gluon distribution of colliding hadron (or nucleus).  We find that the $\jsi$ production is dominated by the color-octet channel, providing an excellent probe to explore the gluon medium in large nuclei at the EIC.
\end{abstract}

\date{\today}

\maketitle
{\it Introduction.---}
 $\jsi$ production has been a focus of theoretical and experimental interest since its discovery over 45 years ago
 \cite{Aubert:1974js,Augustin:1974xw}.
In particular, understanding its production mechanism and using it as a probe to QCD matter are among the most active research subjects in particle and nuclear physics \cite{Brambilla:2010cs}. The challenge is to understand the nonperturbative transition from a produced $\ccb$ pair to a physical $\jsi$.  The non-relativistic QCD (NRQCD) factorization approach \cite{Bodwin:1994jh} to describe $\jsi$ production in hadronic collisions is by far the most phenomenologically successful one, although there are still challenges to understand the polarization of produced $\jsi$ and to resolve the difference between various sets of non-perturbative long distance matrix elements (LDMEs) extracted from the world data \cite{Bodwin:2014gia, Butenschoen:2011yh, Chao:2012iv, Gong:2012ug,Bodwin:2015iua}.  These LDMEs characterize the transition rates for various color-spin states of a produced $\ccb$ pair to become a $\jsi$ and should be process-independent.  The universality of these LDMEs, which has not been confirmed, not even the sign, is a critical test of the NRQCD factorization approach to the $\jsi$ production \cite{Lansberg:2019adr,Chung:2018lyq}.

The $\jsi$ production has been studied in every possible high energy collisions from 
$e^+e^-$ \cite{Pakhlov:2009nj,Abdallah:2003du}, photon-hadron \cite{Chekanov:2002at,Adloff:2002ex,Aaron:2010gz}, 
and hadron-hadron ($hh$) collisions \cite{Acosta:2004yw,Adare:2009js,Abelev:2009qaa,Khachatryan:2015rra,Aad:2015duc,Abelev:2012gx,Aaij:2013yaa}.  
New ideas have been proposed to study $\jsi$ production more differentially, such as measuring a $\jsi$ within a produced jet \cite{Aaij:2017fak,Sirunyan:2019vlp,Bain:2017wvk,Kang:2017yde} as well as its polarization \cite{Kang:2017yde}. With the recent decision to build the Electron-Ion Collider (EIC) \cite{Accardi:2012qut}, in this Letter, we propose to use transverse momentum $p_T$ distribution of single inclusive $\jsi$ production at the EIC to explore and to test the $\jsi$ production mechanism.  Unlike the traditional semi-inclusive deep inelastic scattering (SIDIS), we do not require to measure the scattered electron and define the $p_T$ of produced $\jsi$ in the electron-hadron frame.  With a single observed hard scale, $p_T$, we apply QCD and QED collinear factorization to colliding hadron and electron, respectively, and NRQCD factorization to the hadronization of produced $\ccb$ pair to a physical $\jsi$.  The inclusiveness from not measuring the scattered electron helps us to eliminate a major uncertainty of QED rediative corrections in SIDIS \cite{Liuetal:2020}. We also discuss the similarity and difference from tagging or not tagging the scattered electron.  

In this letter, we perform explicit calculations of short-distance hard parts of $\jsi$ production at both leading order (LO) and next-to-leading order (NLO) in strong coupling constant $\alpha_s$.  We use the recently developed dipole subtraction method \cite{Butenschoen:2019lef} to deal with the QCD divergences for heavy quarkonium production. With its high design luminosity \cite{Accardi:2012qut}, we find from our NLO results that $\jsi$ can be produced at a wide range of $p_T$ with a large number of events at the EIC.  Most importantly, we find that the single inclusive $\jsi$ production at the EIC is very sensitive to the difference between all available sets of LDMEs, which could provide new insights into the $\jsi$ production mechanism. In addition, we are confident to predict that the $\jsi$ produced at the EIC is likely unpolarized, and the production is dominated by gluon initiated partonic subprocesses.  With the dominance of producing color-octet $\ccb$ pairs, the $p_T$-distribution of $\jsi$ production at the EIC could be an excellent probe to explore the dense gluon medium in large nuclei, which should further motivate the measurement of single inclusive $\jsi$ production at the EIC.

{\it Factorization and NLO calculation.---}
We consider single inclusive $\jsi$ production in electron-hadron ($eh$) collisions, 
$e + h\to \jsi(p) + X, $
where the scattered outgoing electron is unobserved/integrated over. 
Single inclusive jet production in $eh$ collisions without observing the outgoing electron was introduced in Ref.~\cite{Kang:2011jw}. Additional studies of the same process were carried out at NLO  \cite{Hinderer:2015hra, Hinderer:2017ntk, Boughezal:2018azh} and even at NNLO \cite{Abelof:2016pby} in recent years. 
A key difference between the jet production and $\jsi$ production at high $p_T$ is that a $\jsi$ is identified in the final-state, and its non-perturbative hadronization process might be interfered by the presence of soft gluons from the beam jet of the colliding hadron.  Following the arguments given for QCD collinear factorization at both leading power and next-to-leading power production of $\jsi$ at high $p_T$ in hadron-hadron collisions \cite{Nayak:2005rt,Nayak:2006fm,Kang:2014tta,Kang:2014pya}, we are confident that the impact of such soft gluon interactions should be suppressed by powers of $\Lambda_{\rm QCD}/p_T$.  Although there is no all-order proof of NRQCD factorization to represent the hadronization of a $\ccb$ pair to a $\jsi$ by an expansion of local LDMEs organized by powers of $\alpha_s$ and the pair's relative velocity, the factorization was found valid up to two-loop calculations \cite{Nayak:2005rt,Nayak:2006fm}.  In this Letter, as an Ansatz, we apply NRQCD factorization for the $\jsi$ hadronization. With a single hard scale $p_T$, the same factorization formalism for $\jsi$ production in hadron-hadron collisions should also apply for electron-hadron collisions,
\bea
\label{eq:fac}
d\sigma_{eh\to\jsi(p)} =& \sum_{a,b,n} f_{a/e}(x_a,\mu_f^2) \otimes f_{b/h}(x_b,\mu_f^2) 
\\
& \otimes \hat{\sigma}_{ab \to c\bar c[n] }(x_a,x_b,p_T,\eta,m_c,\mu_f^2) \mopn
\nonumber
\eea
with $\jsi$ rapidity $\eta$ and factorization scale $\mu_f$. In Eq.~(\ref{eq:fac}), $a=e,\gamma$ in our calculations and could include all parton flavors at higher orders; $b=q,\bar{q},g$; $n$ sums over all color-spin states of the $\ccb$ pair; $\otimes$ represents the convolution over momentum fractions $x_a$ and $x_b$; $f_{a/e}$ is a collinear distribution to find an electron, a photon or a parton from the colliding electron; $f_{b/h}$ is the parton distribution function (PDF) of the colliding hadron; $\hat{\sigma}$ is the perturbatively calculable short distance hard part; and $\mopn$ is the local LDME representing the probability for the state $\ccb[n]$ to become a $\jsi$. With the available collision energies at the EIC, we did not include the fragmentation contribution from resummation of $\log(p_T/m_c)$-type logarithmic contributions \cite{Nayak:2005rt,Nayak:2006fm,Kang:2014tta,Kang:2014pya}.

\bef
\psfig{file=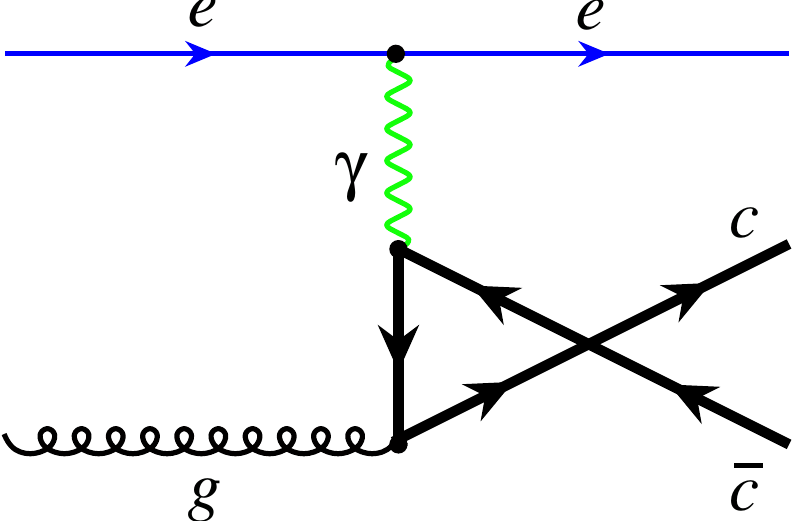, width=1.1in}
{\hskip 0.1in}
\psfig{file=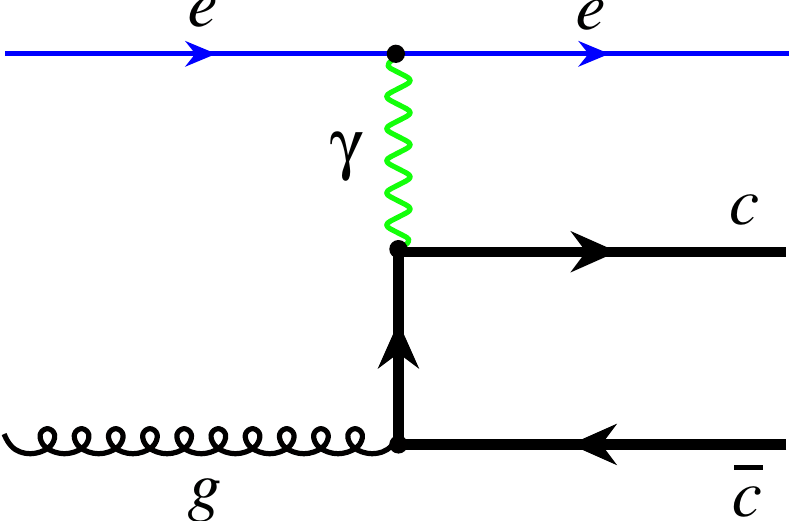, width=1.1in}
\caption{Leading order Feynman diagrams for $\jsi$ production in $eh$ collisions.}
\label{fig-lo}
\eef

We calculate the $\hat{\sigma}$ in Eq.~(\ref{eq:fac}) at both LO and NLO in strong coupling $\alpha_s$, together with leading contributions in electromagnetic coupling $\alpha$, such that 
$
d\sigma_{eh\to\jsi(p)} \approx d\sigma^{LO} + d\sigma^{NLO}.
$
At the LO, as shown in Fig.~\ref{fig-lo}, only gluon channel contributes, 
\bea
d\sigma^{LO} = \sum_{n=\COaSz,\COcPj} f_{g/p}
\otimes\hat{\sigma}^{(2,1)}_{e+g \to c\bar c[n]} \mopn,
\eea
where leading order $f_{e/e}=\delta(1-x_a)$ was used, $\hat{\sigma}^{(2,1)}$ denotes partonic cross section evaluated at $\mathcal O(\alpha^2\alpha_s)$.  At this order, only color octet $\COaSz$ and $\COcPj$ $\ccb$ pair can contribute to high $p_T$ $\jsi$ production in $eh$ collisions.  On the other hand, all four leading $\ccb$ states ($\CScSa$, $\COaSz$, $\COcSa$, $\COcPj$) contribute at the LO in $hh$ collisions.  That is, it is advantageous for $eh$ collisions to get better information on $\mopa$ and $\mopc$ \footnote {For $\COcPj$ channel, we converted all P-wave LDMEs according to $\mopj \approx (2J+1)\mopc$.}. 
 
\bef
\psfig{file=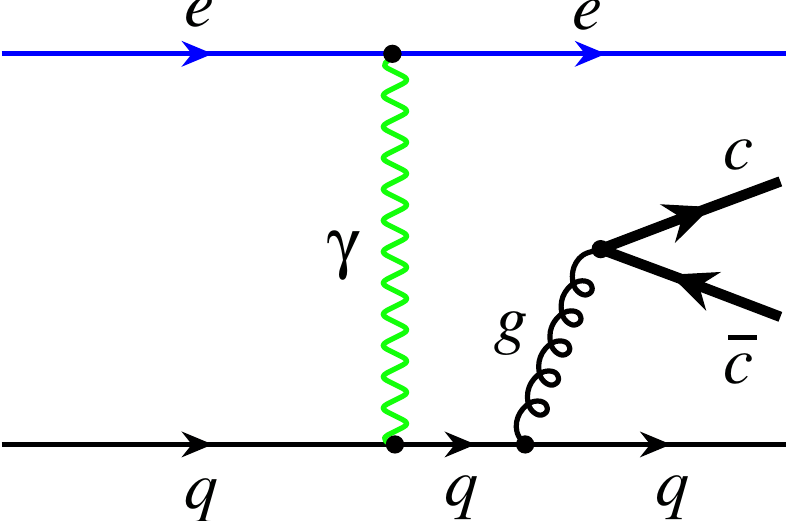, width=0.75in} {\hskip 0.01in}
\psfig{file=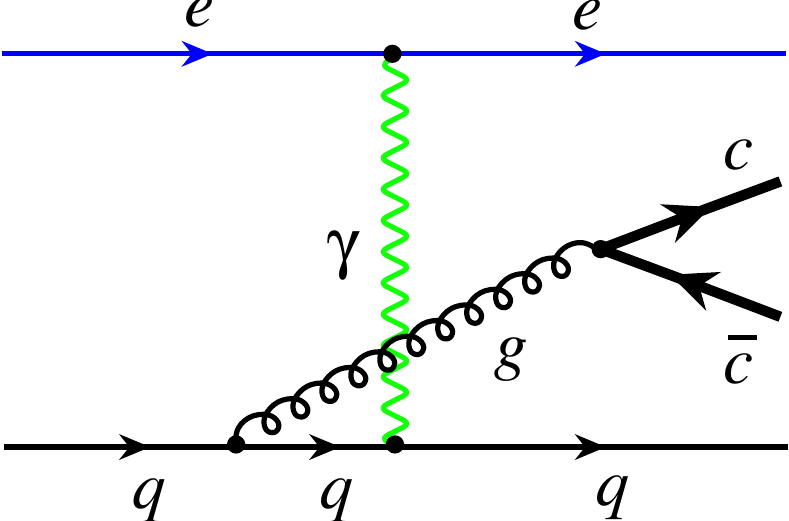, width=0.75in} {\hskip 0.01in}
\psfig{file=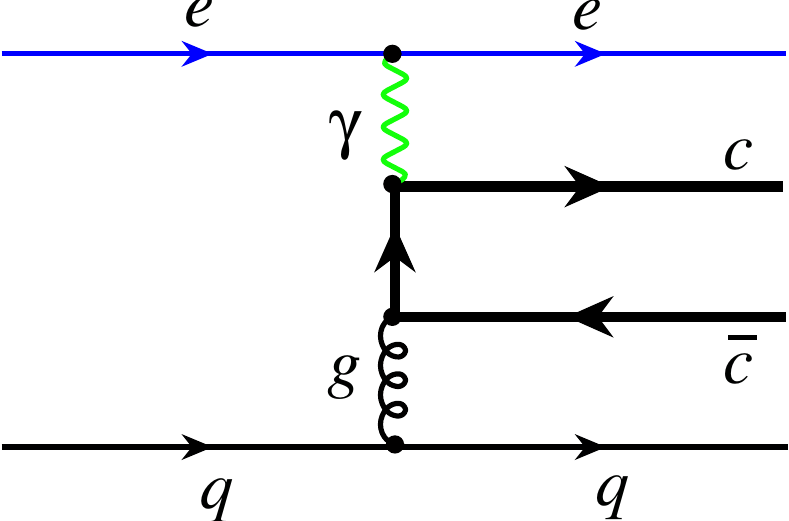, width=0.75in} {\hskip 0.01in}
\psfig{file=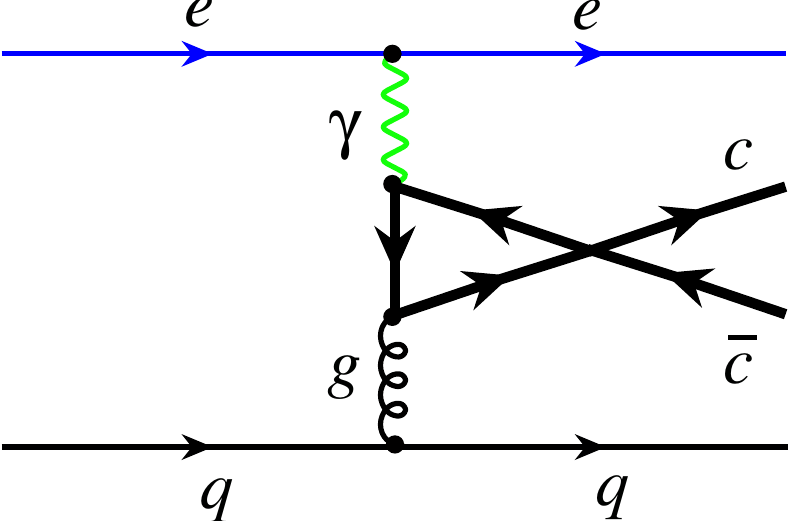, width=0.75in}\\
\vspace{0.15in}
\psfig{file=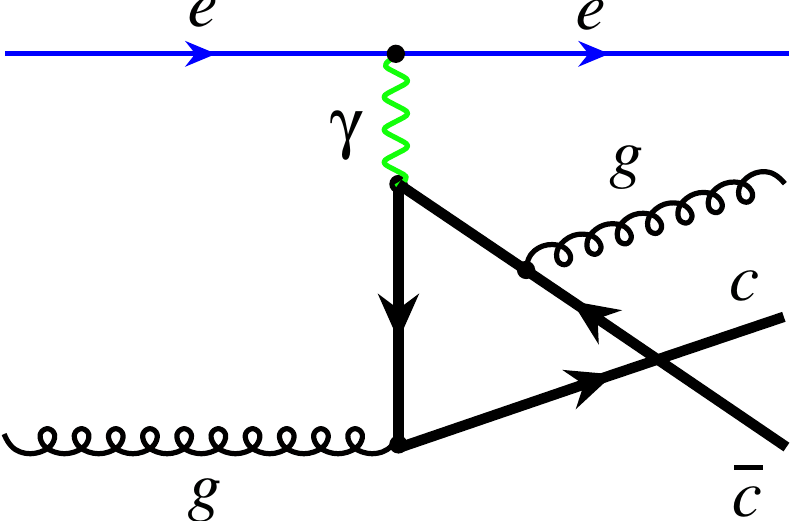, width=0.75in} {\hskip 0.01in}
\psfig{file=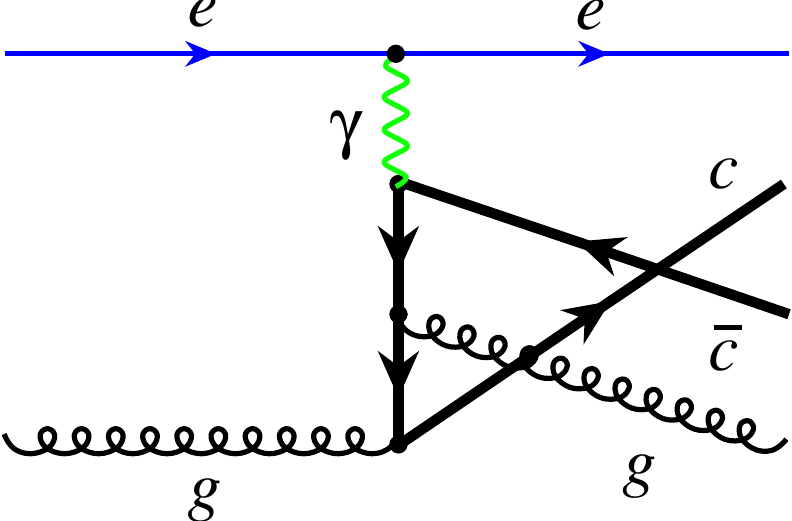, width=0.75in} {\hskip 0.01in}
\psfig{file=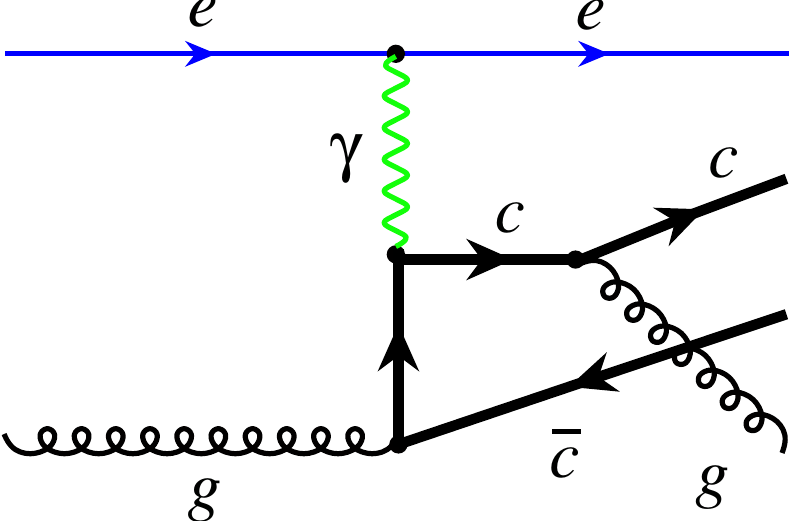, width=0.75in} {\hskip 0.01in}
\psfig{file=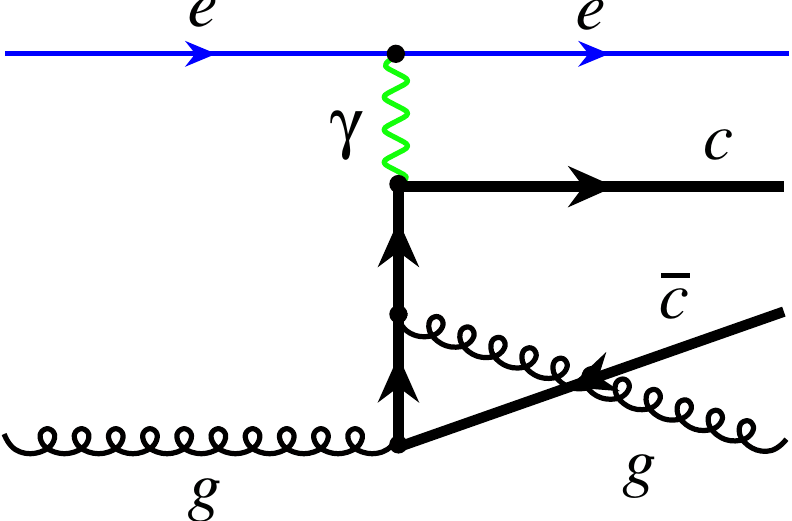, width=0.75in}\\
\vspace{0.15in}
\psfig{file=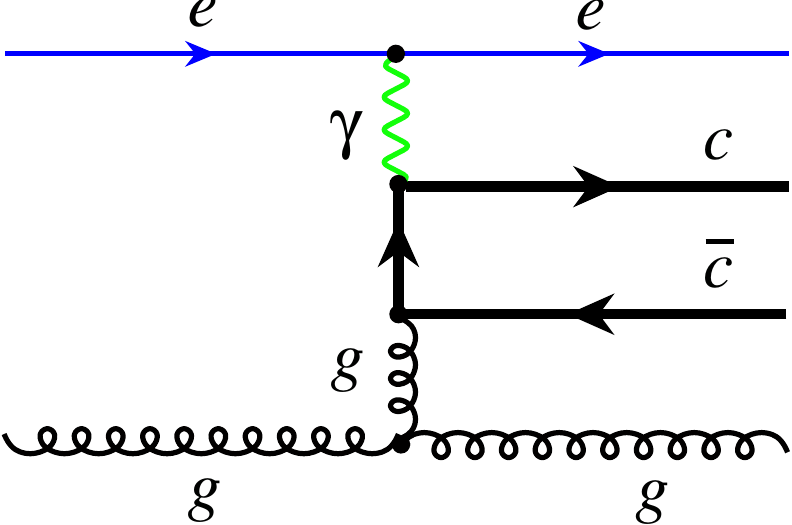, width=0.75in} {\hskip 0.01in}
\psfig{file=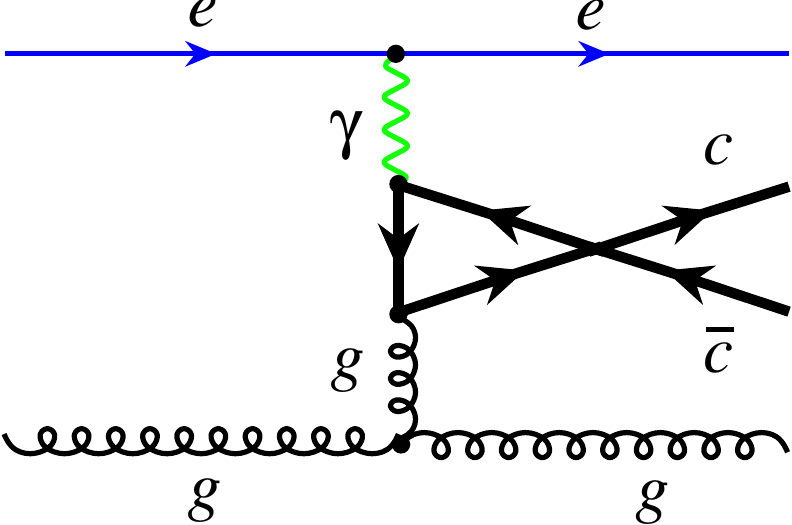, width=0.75in} {\hskip 0.01in}
\psfig{file=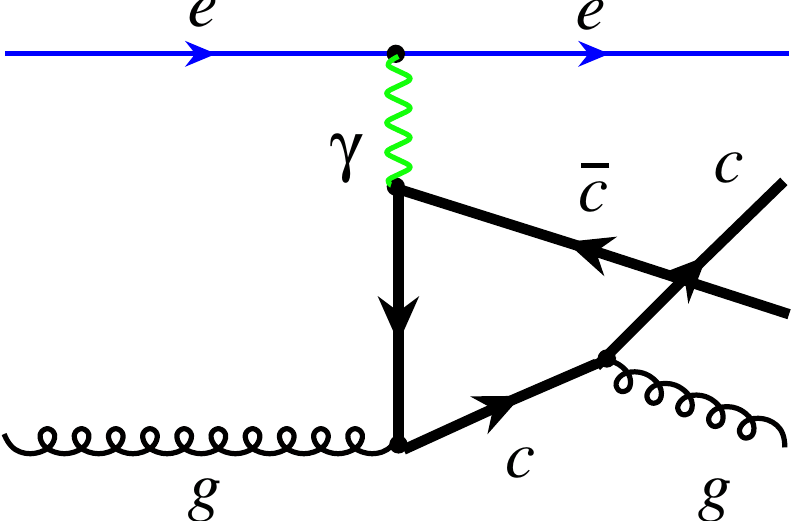, width=0.75in} {\hskip 0.01in}
\psfig{file=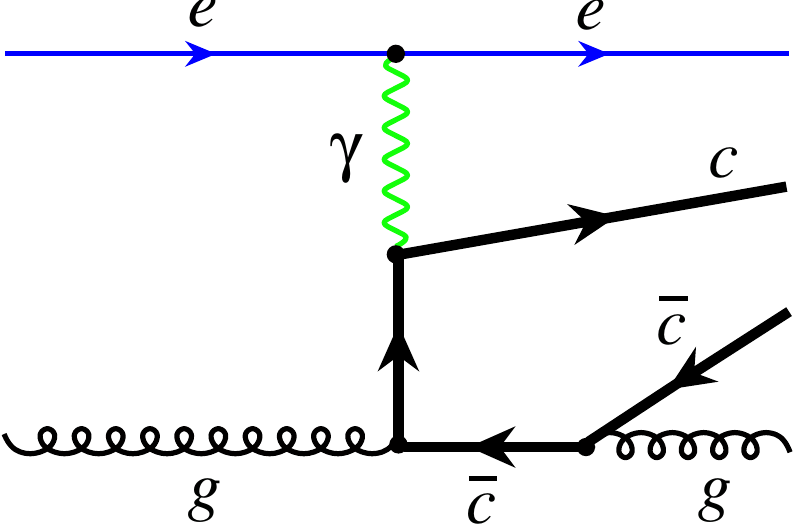, width=0.75in} 
\caption{NLO Feynman diagrams for real contribution to $\jsi$ production in $eh$ collisions.}
\label{fig-real}
\eef

At the NLO, one has to include both real and virtual contributions. For real contribution, new light-flavor quark (antiquark) channels open up, in addition to the gluon channel, as shown in Fig.~\ref{fig-real}.  Since we do not measure the outgoing electron, the real contribution has QED collinear (CO) divergence (if the electron mass $m_e\to 0$) when the outgoing electron is collinear to the incoming one.  Like QCD factorization, this QED CO divergence should be absorbed into $f_{\gamma/e}$. We adopt the method presented in Ref.~\cite{Dittmaier:2008md} for QED divergence. In addition, there are QCD divergences.  The outgoing light parton in Fig.~\ref{fig-real} could be either soft or collinear to the incoming parton. The CO divergence should be absorbed into the PDFs of colliding hadron and the infrared (IR) divergence should be either canceled by the virtual diagrams in Fig.~\ref{fig-virtual} or by soft one-loop correction to LDMEs.  In practice, we use the dipole subtraction method, recently developed specifically for heavy quarkonium production \cite{Butenschoen:2019lef}. This method is based on the standard dipole subtraction techniques for light and heavy flavors \cite{Catani:1996vz, Phaf:2001gc} and has been successfully implemented into calculations for $hh$ collisions \cite{Butenschoen:2020mzi}. 

\bef
\psfig{file=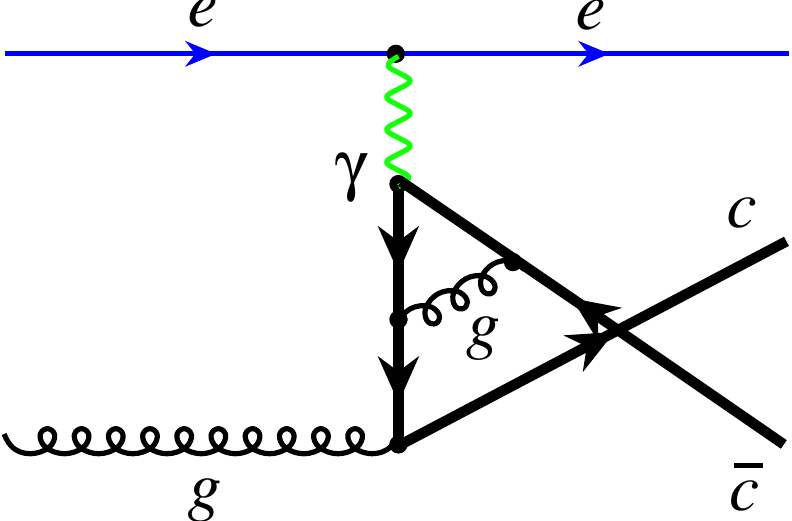, width=0.75in} {\hskip 0.01in}
\psfig{file=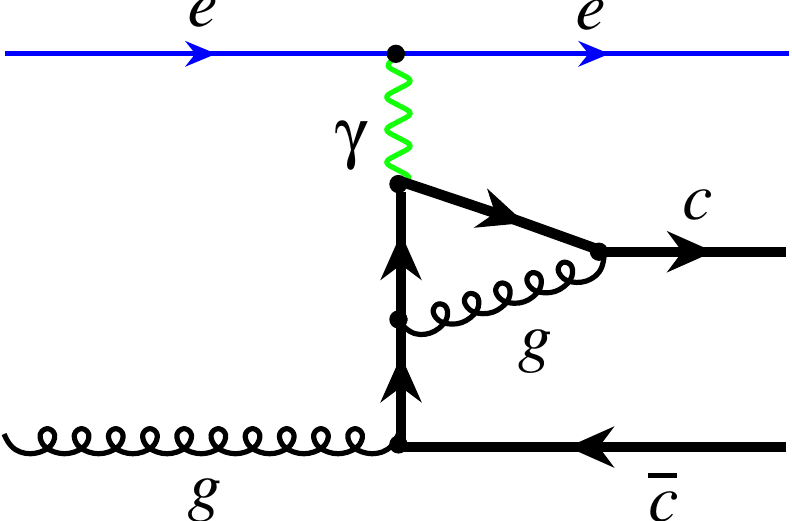, width=0.75in} {\hskip 0.01in}
\psfig{file=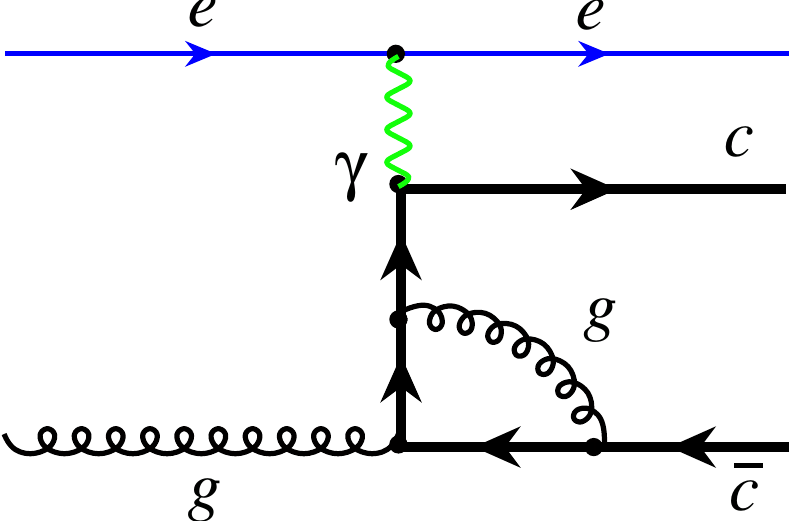, width=0.75in} {\hskip 0.01in}
\psfig{file=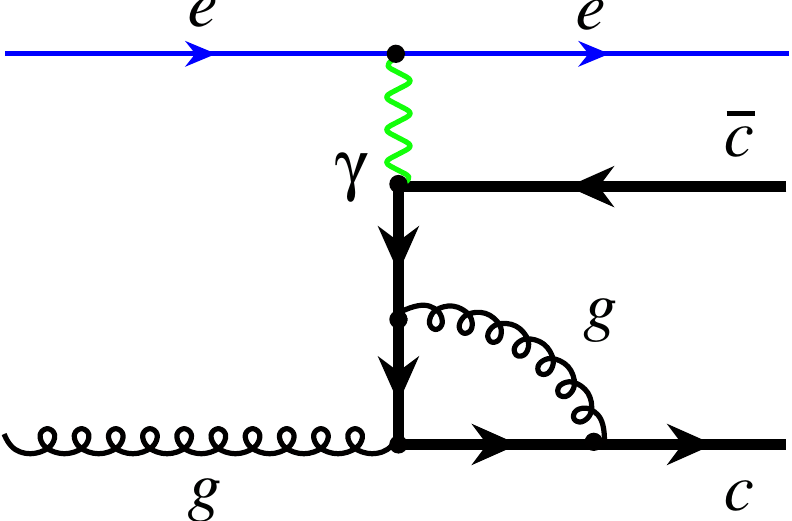, width=0.75in}\\
\vspace{0.15in}
\psfig{file=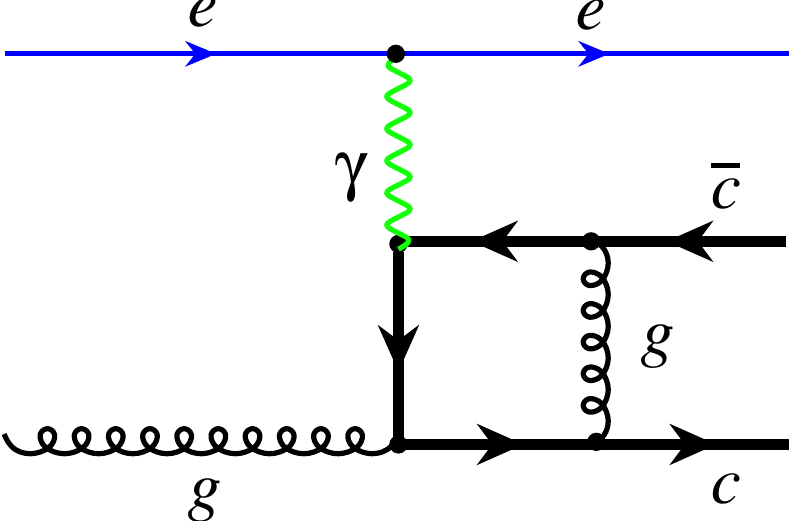, width=0.75in} {\hskip 0.01in}
\psfig{file=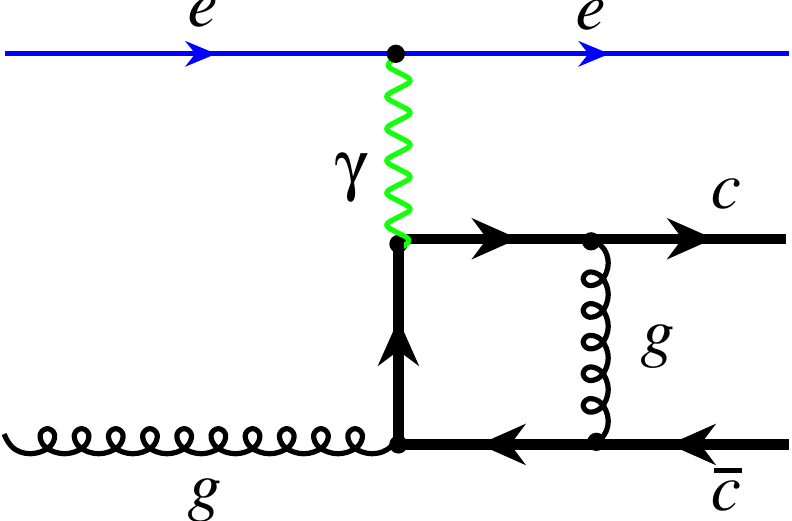, width=0.75in} {\hskip 0.01in}
\psfig{file=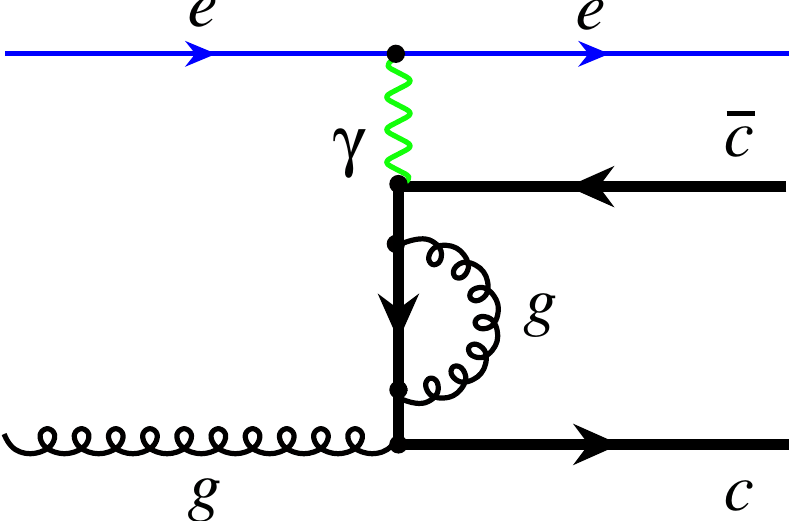, width=0.75in} {\hskip 0.01in}
\psfig{file=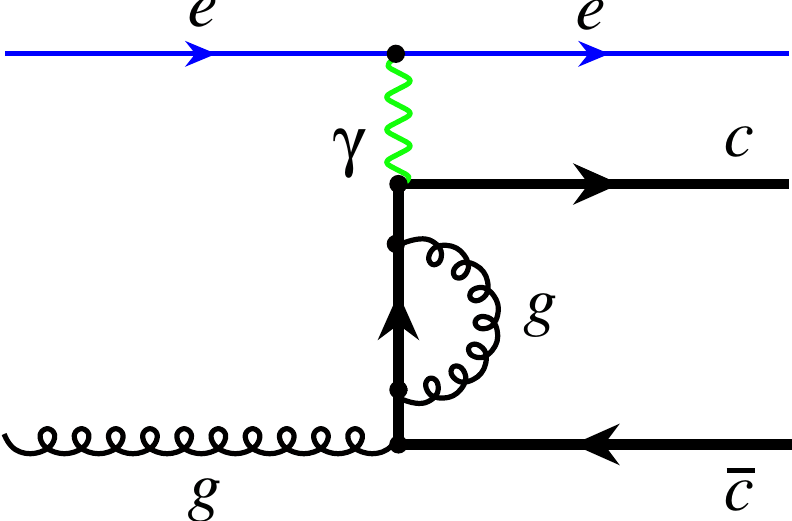, width=0.75in} \\
\vspace{0.15in}
\psfig{file=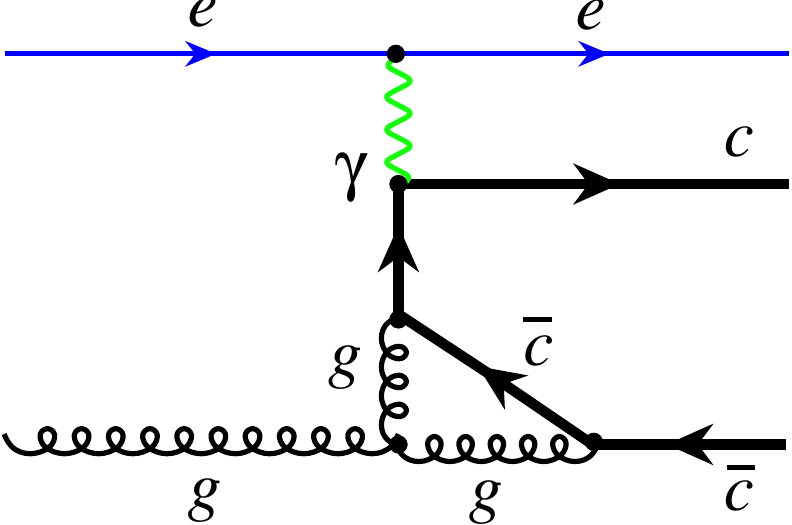, width=0.75in} {\hskip 0.01in}
\psfig{file=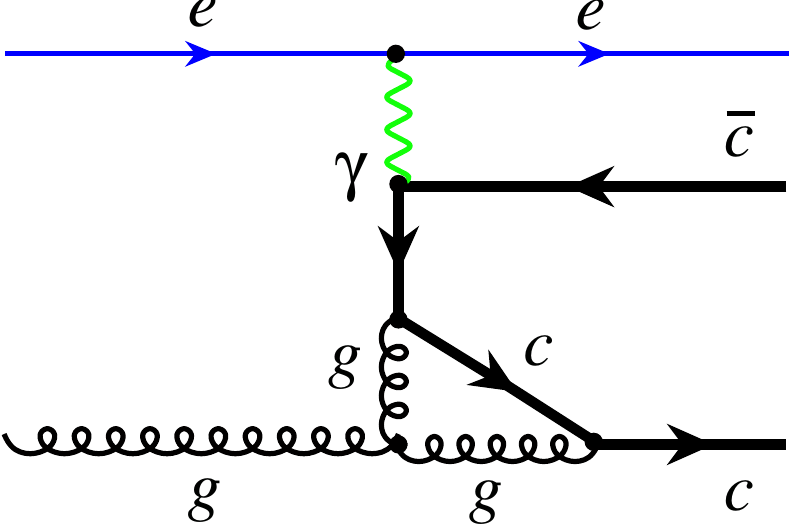, width=0.75in} {\hskip 0.01in}
\psfig{file=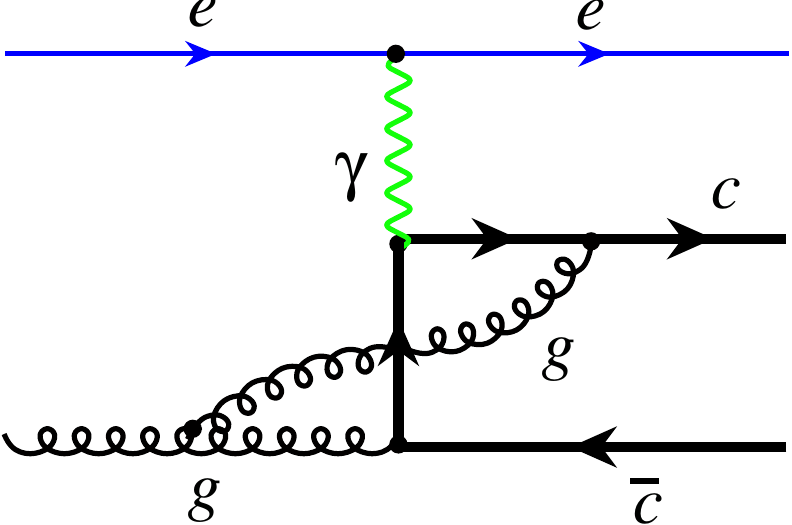, width=0.75in} 
\caption{NLO Feynman diagrams for virtual contribution to $\jsi$ production in $eh$ collisions.}
\label{fig-virtual}
\eef

For the virtual contribution, shown in Fig.~\ref{fig-virtual}, the ultraviolet (UV) and IR divergences can be extracted through evaluation of some standard one-loop scalar integrals. Combining real and virtual contributions together, we obtain the finite partonic hard parts at NLO, 
\bea
d\sigma^{NLO} =& \sum_{b,n} 
\Big[\hat{\sigma}^{(2,2)}_{eb \to c\bar c[n] } 
+  f_{\gamma/e} \otimes \hat{\sigma}^{(1,2)}_{\gamma b \to c\bar c[n] } \Big] 
\otimes f_{b/h} 
\nnu
&\times \mopn ,
\label{eq-xsec_exp}
\eea
where $\sum_n$ runs over four $\ccb$ states:  $\CScSa$, $\COaSz$, $\COcSa$, and $\COcPj$.
The $\hat\sigma^{(2,2)}$, $\hat\sigma^{(1,2)}$ represent the finite contributions at $\mathcal O(\alpha^2\alpha_s^2)$ and $\mathcal O(\alpha\alpha_s^2)$, respectively. Notice that the QED CO divergence has been removed in $\hat{\sigma}^{(2,2)}$, the $\hat\sigma^{(1,2)}$ in Eq.~(\ref{eq-xsec_exp}) corresponds to the region where the exchanged photon is quasi-real. In the leading logarithmic approximation, the leading order photon distribution in an electron is given by
\bea
f_{\gamma/e}(x,\mu_f^2) = \frac{\alpha}{2\pi} \frac{1+(1-x)^2}{x} \left[\ln\frac{\mu_f^2}{x^2m_e^2}-1 \right],
\label{eq-photon}
\eea
with factorization scale $\mu_f$.  An all order resummation of the logarithms could be carried out by solving a QED version of DGLAP evolution \cite{Liuetal:2020}. In this Letter, we skip all details of standard NLO calculation.  Our partonic hard parts, $\hat{\sigma}^{(2,1)}$, $\hat{\sigma}^{(2,2)}$ and $\hat{\sigma}^{(1,2)}$ are used for the following numerical results. 
 
{\it Phenomenology for EIC.---}
We perform following numerical analysis for the EIC kinematics, and our evaluated cross section is fully differential, which gives us the flexibility to implement any kinematic cuts for final state particles including the outgoing electron. We choose a center-of-mass energy at $\sqrt{s} = 141.4$~GeV, $\jsi$ pseudo-rapidity cut $|\eta|<4$, and $\jsi$ transverse momentum cut $3 <p_T < 15$~GeV to be comfortable with QCD factorization and enough $\jsi$ events.  We use CT14-nlo \cite{Dulat:2015mca} for unpolarized proton PDFs. We choose the factorization and renormalization scales $\mu_r = \mu_f = \sqrt{p_T^2+M^2}$ with $M=3.1$ GeV - the $\jsi$ mass. We include four leading $\ccb$ states for $\jsi$ production with corresponding LDMEs from four groups:  Bodwin {\it et.al.} \cite{Bodwin:2014gia}, Butenschoen {\it et.al.} \cite{Butenschoen:2011yh}, Chao {\it et.al.} \cite{Chao:2012iv}, and Gong {\it et.al.} \cite{Gong:2012ug}, which are referred as Bodwin, Butenschoen, Chao and Gong, respectively, and presented in Table~\ref{table:nrqcd}. Although these LDMEs were extracted from fitting similar data sets, their numerical values are very different, even different in sign in some cases.  

 \begin{footnotesize}
\begin{table}[!t]
\caption{$\jsi$ NRQCD LDMEs from four different groups.}
\label{table:nrqcd}
\begin{center}
  \begin{tabular}{  l |c | c | c | c }
  \hline
    & $\langle {\mathcal O}(^3S_1^{[1]})\rangle$ & $\langle {\mathcal O}(^1S_0^{[8]})\rangle$ & $\langle {\mathcal O}(^3S_1^{[8]})\rangle$ & $\langle {\mathcal O}(^3P_0^{[8]})\rangle$ \\ 
    & GeV$^3$ & $10^{-2}$ GeV$^3$ & $10^{-2}$ GeV$^3$ & $10^{-2}$ GeV$^5$ 
        \\ \hline
    Bodwin & 0 & 9.9 & 1.1  & 1.1 \\ \hline
       Butenschoen & 1.32 & 3.04 & 0.16 & $-0.91$\\ \hline
   Chao & 1.16 & 8.9 & 0.30  & 1.26 \\ \hline
   Gong & 1.16 & 9.7 & $-0.46$ & $-2.14$ \\ \hline
  \end{tabular}
\end{center}
\end{table}
  \end{footnotesize}

\bef
\psfig{file=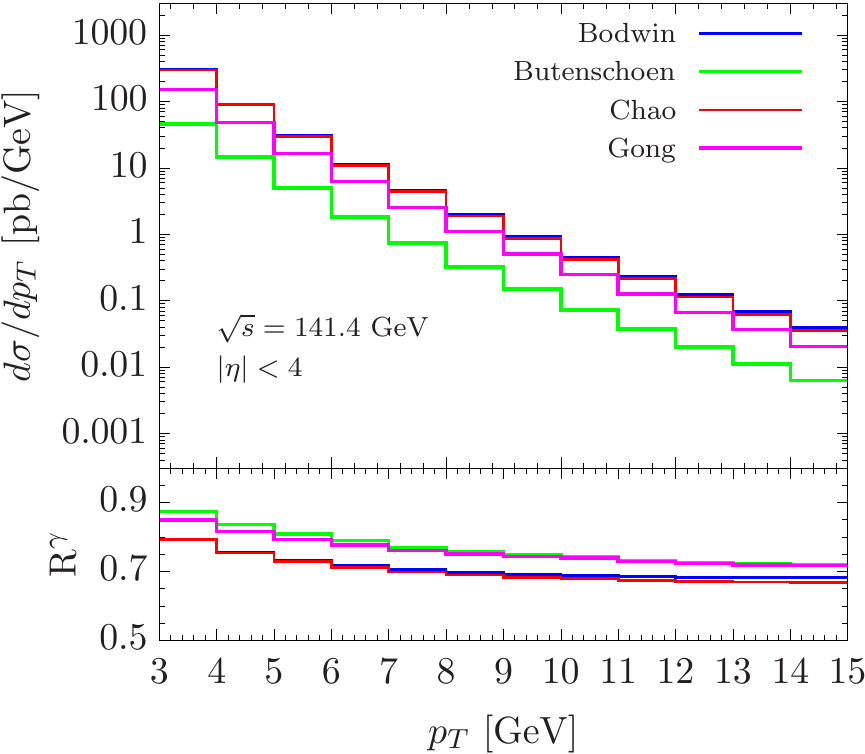, width=3.0in}
\caption{The NLO differential cross section as a function of $p_T$. Four histograms correspond to four different sets of LDMEs. The lower panel shows the fraction from the quasi-real photon channel.}
\label{fig-xsec_pt}
\eef

In order to illustrate the sensitivity to the nonperturbative NRQCD LDMEs, we show in Fig.~\ref{fig-xsec_pt} the differential cross section of inclusive $\jsi$ production, without tagging the outgoing electron, as a function of $p_T$.  It is clear that the production rate, evaluated with all four sets of LDMEs in Table~\ref{table:nrqcd}, is large enough for producing sufficient $\jsi$ events at the future EIC.  It is important to notice that predictions using LDMEs from Bodwin and Chao are very similar, but significantly different from those using other two sets of LDMEs.  In addition, there is almost an order of magnitude difference in production rate between Bodwin/Chao and Butenschoen. Such drastic differences between predictions using the four different sets of LDMEs clearly demonstrates the discriminative power of this new observable on $\jsi$ production mechanism.  The lower panel in Fig.~\ref{fig-xsec_pt} shows the fractional contribution from the channel with a quasi-real photon, $R^{\gamma}$, defined as a ratio of contribution from the quasi-real photon channel ($\hat\sigma^{(1,2)}$ term in Eq.~(\ref{eq-xsec_exp})) and the total contribution.  Clearly, the quasi-real photon channel is dominant for all four sets of LDMEs, especially at lower $p_T$ due to the $1/x$-dependence of $f_{\gamma/e}$ in Eq.~(\ref{eq-photon}). 
Notice that we used the leading logarithmic approximation for the photon distribution in an electron as shown in Eq.~(\ref{eq-photon}), the 70$\%$ contribution from quasi-real photon channel should be reduced once we consider the resummed photon distribution in an electron, but not in a significant way. Detailed analysis considering QED evolution of photon distribution in an electron will be presented in a future publication.

\bef
\psfig{file=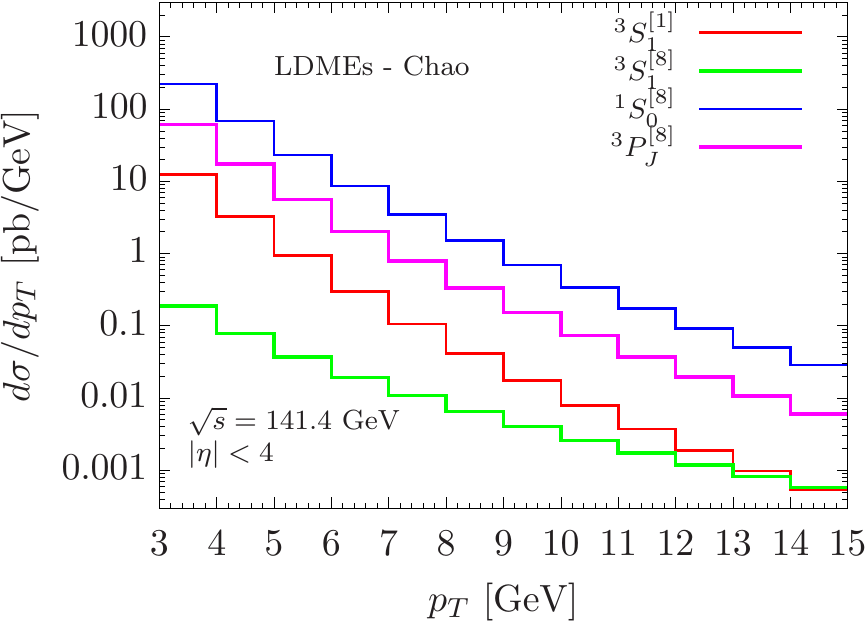, width=3.0in}
\caption{The NLO differential cross section as a function of $p_T$ for $\jsi$ production in $eh$ collisions. The four histograms correspond to the contributions from four different $\ccb$ states, respectively. }
\label{fig-xsec_n}
\eef
To further explore the sensitivity to each particular $\mopn$, we studied the cross section more differentially in terms of four different $\ccb$ states. In Fig.~\ref{fig-xsec_n}, we used the LDMEs from Chao as an example, and found that contribution to the $\jsi$ production is dominated by the production of $\COaSz$ state for whole $p_T$ region.  We checked that this feature also holds true for the other three sets of LDMEs.  The fact that the $\COaSz$ channel clearly dominate the production rate and Bodwin and Chao have a very close value for corresponding $\mopa$ should be the reason for the very similar predictions from Bodwin and Chao as shown in Fig.~\ref{fig-xsec_pt}.  Notice that the extracted $\mopa$ from Gong has even closer value comparing to Bodwin, however, the opposite sign for $\mopc$ in these two sets eventually lead to big difference between the blue and purple histograms in Fig.~\ref{fig-xsec_pt}. The detailed analysis shown in Fig.~\ref{fig-xsec_n} indicates that $\jsi$ production in $eh$ collisions can be served as a very sensitive observable to constrain the LDME $\mopa$. Since $\COaSz$ has no polarization preference, we predict that high-$p_T$ $\jsi$ produced in inclusive $eh$ collisions will be likely unpolarized.  The measurement of $\jsi$ polarization at the future EIC will be a stringent test of NRQCD factorization. 

If we extend this study to electron-nucleus collisions at the EIC, the dominance of production rate for the $\COaSz$ $c\bar c$ state in initial hard production will provide us a unique channel to study how a color octet $\ccb$ state interacts with nuclear medium when it propagates through a large nucleus.

\bef
\psfig{file=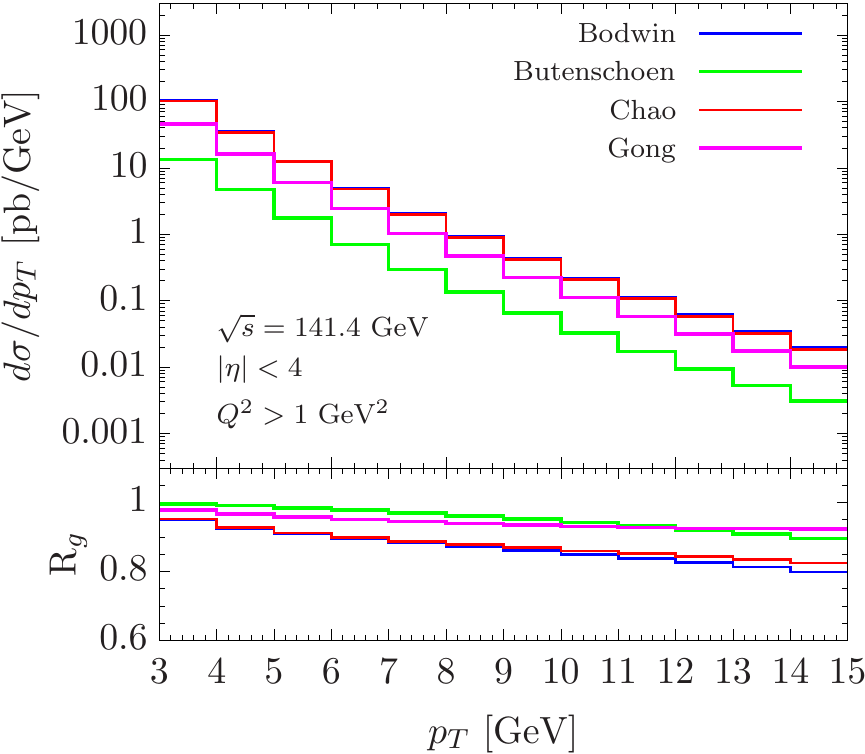, width=3.0in}
\caption{Same as Fig. \ref{fig-xsec_pt}, but with $Q^2>1$ GeV$^2$. The lower panel shows the fraction from initial gluon channel.}
\label{fig-xsec_Qcut}
\eef
Since our result on $\jsi$ production is fully differential in phase space for final state electron, this allows us to study the production of $\jsi$ by exchanging a virtual photon. We impose a constraint on the momentum transfer $Q^2 > 1$ GeV$^2$ for the exchanged photon, so that we could neglect the contribution from quasi-real photon channel. In this case, the QED CO divergence will be regularized by $Q^2$ cut, and the QED dipole subtraction in $\hat\sigma^{(2,2)}$ has to be removed. In Fig.~\ref{fig-xsec_Qcut}, we show the $\jsi$ cross section with $Q^2$ cut. As expected, the production rate becomes smaller compare to the result without constraint on $Q^2$, but still different sets of LDMEs lead to very different production rate.  We further isolate the contributions from initial quark and gluon channels. In the bottom panel of Fig.~\ref{fig-xsec_Qcut}, we plot $R_g$ as the gluon initiated fraction of total differential cross section.  The large value of $R_g$ in a wide $p_T$ region indicates that $\jsi$ production in $eh$ collisions is dominated by initial gluon channel.  This feature makes $\jsi$ production in $eh$ collisions a good observable to probe the initial gluon distribution in colliding proton or nucleus at the EIC.  Understanding the glue is one of the main science goals for future EIC.

{\it Summary.---}
We proposed to measure the $p_T$ distribution of inclusive $\jsi$ production in the electron-hadron frame at the future EIC without tagging the outgoing electron.  We applied QCD and QED collinear factorization to the production of a $\ccb$ pair at high $p_T$, and non-relativistic QCD factorization to the hadronization of the pair to a $\jsi$, and argued that such factorization should be consistent with the factorization formalism used for hadron-hadron collisions.  
Using the dipole subtraction method within the framework of NRQCD, we performed explicit calculations at both LO and NLO in $\alpha_s$.  We found that the existing four sets of NRQCD LDMEs give very different predictions for this new proposed observable, which clearly demonstrated the uniqueness of future EIC in studying the $\jsi$ production mechanism. We also found that the dominance of producing the $\COaSz$ $\ccb$ state in the total contribution provides a solid prediction that $\jsi$ produced in $eh$ collisions will likely be unpolarized. This prediction can provide a stringent test of NRQCD factorization and shed a new light on the $\jsi$ production mechanism. Without tagging the outgoing electron, this observable will not be sensitive to the major uncertainty from QED radiative corrections in the traditional SIDIS. In addition, this new observable at the EIC could provide even more opportunities from its sensitivity to initial-state gluon distribution in nucleon or nucleus and its dominance to produce a color octet $\ccb$ state propagating through the nuclear medium in $eA$ collisions.   All these unique features of this new observable will make the $p_T$ distribution of single inclusive $\jsi$ production in $eh$ collisions a new and very important channel to study at the future EIC.

This work of J.-W.Q. is supported by the U.S. Department of Energy contract DE-AC05-06OR23177, under which Jefferson Science Associates, LLC, manages and operates Jefferson Lab.
The work of X.-P. W. is supported by the U.S. Department of Energy, Division of High Energy Physics, under Contract No. DE-AC02-06CH11357. The work of H. X. is supported by the research startup funding at South China Normal University. 
This work is also supported within the framework of the TMD Topical Collaboration.

\end{document}